\documentclass[aps,prb,twocolumn]{revtex4-1} 
\usepackage{graphicx} 
\usepackage{gensymb} 
\usepackage{textcomp}

\begin{document}

\title{Composition profiling InAs quantum dots and wetting layers by atom probe tomography and cross-sectional scanning tunnelling microscopy}

\author{A.D. Giddings$^1$}
\author{J.G. Keizer$^2$}
\author{M. Hara$^1$}
\author{G.J. Hamhuis$^2$}
\author{H. Yuasa$^1$}
\author{H. Fukuzawa$^1$}
\author{P.M. Koenraad$^2$}

\affiliation{$^1$Corporate R\&D Center, Toshiba Corporation, 1, Komukai Toshiba-cho, Kawasaki, 212-8582, Japan} 
\affiliation{$^2$Department of Applied Physics, Eindhoven University of Technology, P.O. Box 513, NL-5600 MB, Eindhoven, The Netherlands}

\begin{abstract}
This study compares cross-sectional scanning tunnelling microscopy (XSTM) and atom probe tomography (APT). We use epitaxially grown self-assembled InAs quantum dots (QDs) in GaAs as an exemplary material with which to compare these two nanostructural analysis techniques. We studied the composition of the wetting layer and the QDs, and performed quantitative comparisons of the indium concentration profiles measured by each method. We show that computational models of the wetting layer and the QDs, based on experimental data, are consistent with both analytical approaches. This establishes a link between the two techniques and shows their complimentary behaviour, an advantage which we exploit in order to highlight unique features of the examined QD material.
\end{abstract}




\maketitle

\section{Introduction}

Accurate and high-resolution structural imaging and compositional analysis techniques have been the key driving technological force behind recent advances in nanotechnology and nanoanalysis. Today, a multitude of analysis tools are employed by scientists and engineers studying nanostructures, each offering its own specific capabilities and advantages. Because each technique can supply different data, when a complex material or structure is being considered it is common that several methods will be used together in order to build a more complete and accurate understanding of the subject.

Of the various methods in use, there are two techniques that are of particular interest thanks to their unparalleled ability to provide atomic-level imaging. The first technique, cross-sectional scanning tunnelling microscopy (XSTM), one of the family of scanning probe microscopy techniques, is well suited for studying semiconductor materials, particularly in the III-V \cite{Kubby1996} and II-VI \cite{Wierts2007} arena. The key strength of XSTM is that it can directly visualize the atomic structure of a material, allowing detailed structural analysis.\cite{Ulloa2009,Mikkelsen2004} However, analysis of nanostructures can be difficult due to the lack of contrast between atomic species, the outward relaxation of the surface, and the two-dimensional nature of the technique.
Despite this, the great precision of the measurements makes it feasible to infer details that can not otherwise be directly visualized. For example, statistical analysis and finite element (FE) calculations can be used to 3D create models of the atomic structure.\cite{Davies2002,Bruls2002}
However, this FE approach, in which the outward relaxation of a strained surface is calculated, has an inherent limitation that it is not injective; any measured dataset can be simulated by a range of different input models.\cite{Mlinar2009}

The second technique under consideration is atom probe tomography (APT). This is the latest evolution of the venerable field emission microscope,\cite{muller_atom-probe_1968} whereby a field ion microscope is combined with a spatially resolved time-of-flight mass spectroscope, creating a device known as a three dimensional atom probe (3DAP).\cite{cerezo_application_1988} In 3DAP a needle shaped specimen of material is placed under a large pulsed voltage causing single ions to be emitted from the specimen's apex and accelerated towards a detector screen. The data collected after each voltage pulse can be combined to form a three-dimensional tomographic image of the specimen. This technique works best for materials with a high conductivity, where sub-atomic precision can be achieved, and so traditionally there has been little use for the technique outside of metallic systems.\cite{Larson2004, yuasa_relation_2008} However, the re-emergence of laser based 3DAP has greatly expanded the range of materials that can be analysed.\cite{Thompson2007} Where semiconductor materials were previously out of bounds, now the additional thermal excitation from a pulsed laser makes performing APT on these materials feasible, although with a reduced analytical quality.\cite{Cerezo2007} Consequently, pioneering work to study semiconductor nanostructures, such as quantum dots (QDs), has started to be undertaken.\cite{Muller2008}

In terms of capabilities, APT seems to complement XSTM very well. Where XSTM can image only 2D cross-sections, APT provides a 3D tomographic reconstruction, and where XSTM has limited capabilities to distinguish between chemical species, the mass-spectral analysis offers the ability not only to distinguish between different elements but also different isotopes. Naturally, these capabilities do not come without a price. The length-scales of the volume that can be measured by the 3DAP are typically of the order of tens of nm laterally and hundreds of nm vertically, which is far more restrictive than the many \micro m that can be imaged with XSTM. Furthermore, the spatial precision is not as great as with XSTM, and not all the emitted ions can be detected. Even in ideal conditions detection efficiency is less than 60\%.\cite{Gault2009} Finally, a particular weakness of APT is that the images of the sample must be reconstructed, a process requiring many assumptions about factors such as apex shape, radius, evaporation conditions and so forth.\cite{Miller1996,Miller2000} The result is that, whilst the 3DAP provides a very unique dataset, the reliability and spatial accuracy is inherently inferior to the direct measurements of XSTM.

\section{Experimental Details}

The studied material is comprised of (In$_x$,Ga$_{1-x}$)As self-assembled QD layers in GaAs, grown on an (100) GaAs substrate via molecular beam epitaxy (MBE). The QD layers were grown by deposition of 2~ML of InAs at a rate of 0.11~ML/s and at a substrate temperature of 500$^{\circ}$C. Formation of the QDs was observed by RHEED. Vertically, the structure consists of five InAs QD layers separated by 50~nm of GaAs with a 20~nm cap. The separation between the wetting layers (WLs) is considered sufficient to suppress strain induced nucleation.\cite{Xie1995}

From atomic force microscopy (AFM) measurements on uncapped material grown under the same conditions, the areal density of the QDs was determined to be $\approx 3.0\,\times\,10^{10}$~cm$^{-2}$. Low temperature (5~K) macro photoluminescence measurements show that the QDs emit at an energy of 1.229~eV with a FWHM of 59~meV.

The APT measurements were performed using an LEAP 3000X Si instrument, operating in laser mode. The laser wavelength was green (532~nm) with a pulse frequency of $0.5$~MHz (2~\micro s) and a target evaporation rate was set to be 0.2\%. The flight length between specimen and detector was 90.0~mm and the typical time-of-flight of an evaporated ion was 0.6--1.6~\micro s, depending on the species. The evaporation rate refers to the probability that, for a given laser pulse, an event will be recorded by the detector. In this experiment, therefore, we aimed to achieve a single detection event every 500 pulses, that is, on average there would be 2~ms between each ion strike. The evaporation field is dynamically altered during the measurement in order to maintain this target. In general, the field will gradually increase over the course of the experiment as the radius of the specimen tip becomes larger as the specimen is evaporated down. Because the position sensitive detector in the 3DAP is not able to accurately distinguish multiple impacts it is important that the evaporation rate is low to maintain a high ratio of single ion events.

The laser pulse energy was less than $0.01$~nJ; such a low power was used in order to ensure good data quality.\cite{gault_influence_2010} The sample stage temperature was set to $50$~K. The needle shaped specimen was fabricated via standard focused ion beam (FIB) techniques using a dual-beam FEI Nova 200 NanoLab system.\cite{thompson_minimization_2006,thompson_situ_2007} Prior to FIB sharpening a protective 100-150~nm Ni layer was sputtered on top of the GaAs sample. Inside the FIB system, a further 100-150~nm Pt layer deposited on the Ni using the ion beam deposition capabilities of the FIB.

All XSTM measurements were performed at room temperature under UHV ($p<6\times10^{-11}$~mbar) conditions with an Omicron STM-1, TS2 Scanner. Electrochemically etched tungsten tips were used. The STM was operated in constant current mode on {\it in situ} cleaved (110) surfaces. To ensure that the electronic contribution to the apparent height, as measured by XSTM, is minimized, a high negative bias voltage ($\approx -3$~V) was applied during all measurements resulting in a purely topographic signal \cite{Feenstra1999} allowing the identification of individual In atoms.\cite{Pfister1996}

The strain relaxation, induced by the lattice mismatch between InAs and GaAs, of the WL and the QDs was modelled with the finite element (FE) method. The FE calculations were performed using the MEMS module of COMSOL Multiphysics. To calculate the strain relaxation of the WL and the QDs a 2D and 3D model was used, respectively.

\begin{figure}
\includegraphics[width=8.6cm]{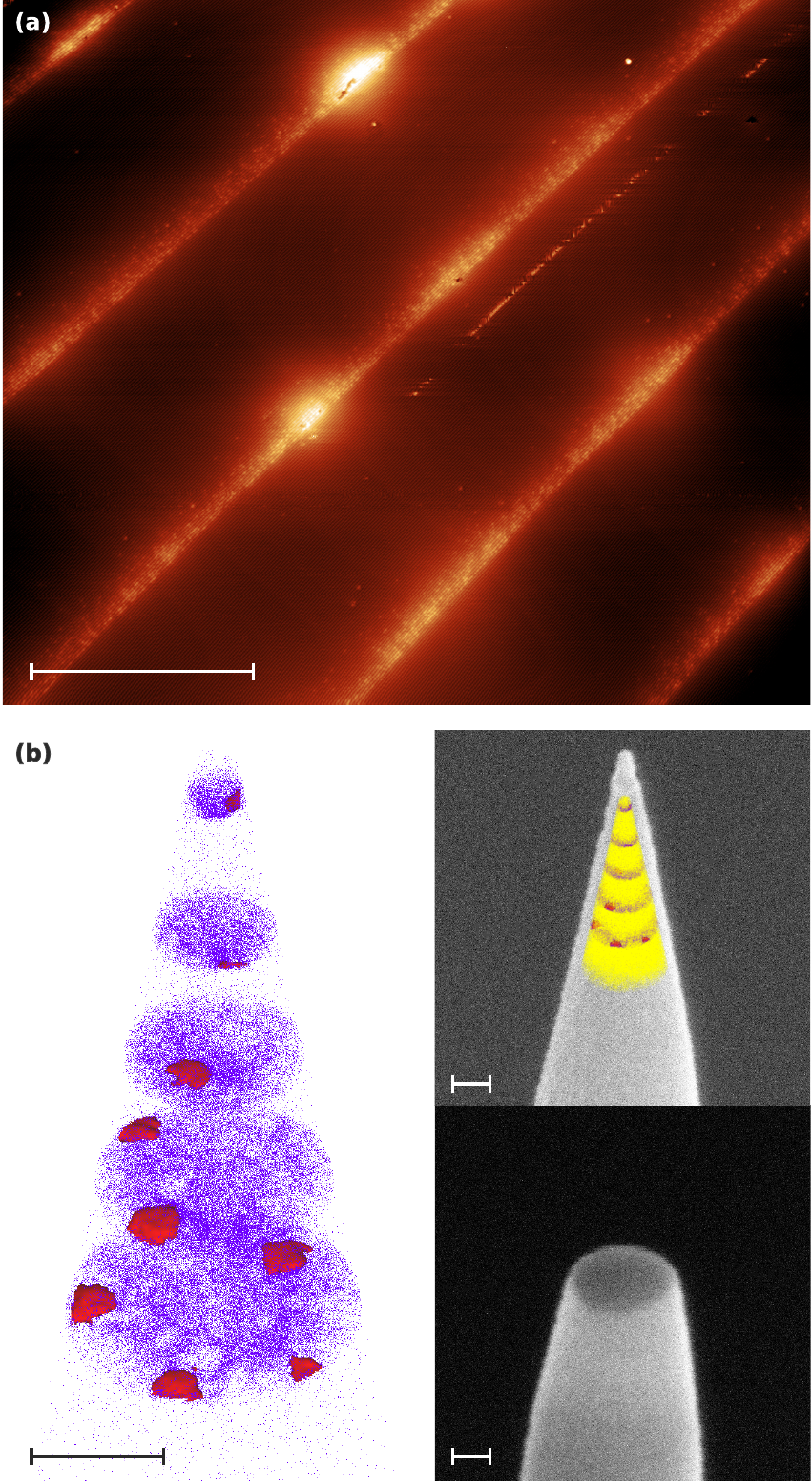}
\caption{Comparative view of typical data obtained with the two techniques under consideration. The five (In$_x$,Ga$_{1-x}$)As layers are clearly visible by both methods. All the scale bars are 50~nm. (a) $190$~nm\,$\times\,165$~nm topographic XSTM image taken at $V=-3$~V, $I=30$~pA. Several QDs can be distinguished as bright features in the WLs. The thin line parallel to the QD layers is an adatom that is being pushed by the tip in the slow scan direction. (b) Atom map showing only the In atoms, marked as indigo pixels. The In atoms between the WL are background noise and account for approximately 0.1--0.6\textperthousand~of the detected atoms. Red isosurfaces show the $x=25\%$ In fraction, marking the location of QDs. The top inset shows the reconstruction, now also with 20\% of the detected Ga atoms (yellow pixels), overlaid on an SEM image of the FIB fabricated specimen prior to evaporation. The bottom inset shows the remains of the specimen after evaporation.}
\label{figure_1}
\end{figure}

\section{Results}

A comparative view of typical data obtained with the two techniques of XSTM and APT is shown in Fig.~\ref{figure_1}. From this figure, it immediately becomes apparent that the two techniques are of a different nature; where the XSTM measurement is restricted to a cleavage plane and thus yields information that is 2D in nature, the APT measurement provides a fully 3D dataset. Even without further analysis, the WLs and QDs can already be distinguished from Fig.~\ref{figure_1}a-b. In the following sections, a detailed comparison is given of the WLs, and the QDs, as measured by XSTM and APT.

\subsection{Wetting Layer}

\begin{figure}
\includegraphics[width=8.6cm]{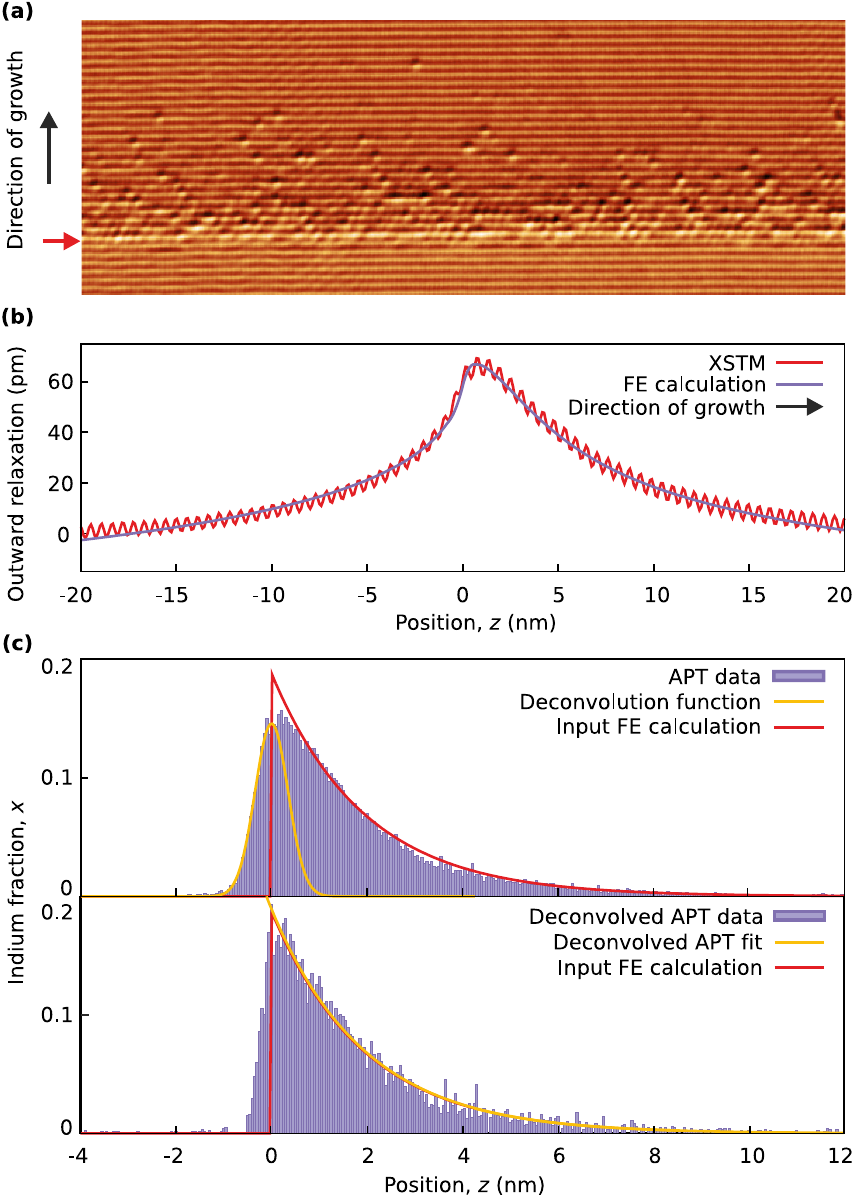}
\caption{WL analysis. (a) $52$~nm\,$\times\,19$~nm local mean equalization filtered current XSTM image of the WL taken at $V=-3$~V, $I=40$~pA. The start of the WL (red arrow) occurs within one bilayer and is followed by a decay of the In concentration in the direction of growth. (b) Outward relaxation profile, averaged over 90~nm of WL, as measured by XSTM (red line) and the result of the FE calculation (blue line). (c)~(top) In fraction of the WL as a function of the position in the direction of growth as measured by APT and the used deconvolution function (yellow line). (c)~(bottom) Deconvolved APT data and the fit of an exponentially decaying function (yellow line), to the data. To ease comparison, the input for the FE calculation is also plotted (red line).}
\label{figure_2}	
\end{figure}

An atomically resolved $52$~nm\,$\times\,19$~nm local mean equalization filtered current XSTM image of a typical part of the WL is shown in Fig.~\ref{figure_2}a. The WL is found to start abruptly (within one bilayer), followed by a decay of the In concentration in the direction of growth. From previous work it is known that the decay of the In concentration can be modelled by an exponentially decaying function.\cite{Offermans2005} The function $a\exp(-z/b)$, with $z=0$ the start of the WL and positive $z$ representing the growth direction, was used as an input for the FE modelling that was employed to calculate the outward relaxation of the cleaved surface due to the strain in the WL. By adjusting the parameters $a$ and $b$ until the calculated outward relaxation profile (blue line, Fig.~\ref{figure_2}b), matches the outward relaxation profile as measured by XSTM (red line, Fig.~\ref{figure_2}b), the decay of the In concentration can be determined. We find $a=0.188\pm0.002$ and $b=1.95\pm0.05$.

The In concentration in the direction of growth can also be determined from the APT data. This is found by sampling the composition of 20~nm diameter cylinders, aligned parallel to the $z$-axis.\footnote{See EPAPS Document No. [number will be inserted by publisher] for a visualization of the WL and sampling cylinder. For more information on EPAPS, see http://www.aip.org/pubservs/epaps.html.}
A small bin length was chosen in order not to loose detail of the sharp concentration changes. The thin diameter was chosen so the cylinder could be positioned not to coincide with any QDs and yet remain close to the core of the specimen, where the reconstruction is most accurate. The thin diameter also helps to reduce the effects of curvature in the reconstruction and misalignment of the cylinder to the normal axis, both of which could result in loss of detail in the sharp onset of the WL. In order to ensure that there were no systematic errors in the reconstructions, volumes from different layers and different specimens were compared. To improve the signal-to-noise ratio, the concentrations of these different sampling volumes were averaged together; this averaged data is shown in Fig.~\ref{figure_2}c~(top).

From the XSTM measurements we unambiguously determined that the onset of the WL is abrupt (within one bilayer). This is not the case in the APT data, where the In fraction rises from $x=2\%$ to 15\% over a distance of $\Delta z \approx0.7$~nm. Despite the measures taken to ensure a good interface, we consider this artifact to arise from the effect of averaging several volumes, which also do not individually have completely sharp interfaces, causing some error in determination of the start of the WL. However, since we know that the WL starts abruptly, we can perform a deconvolution of the APT profile in order to compensate for this. The deconvolution function was determined by fitting a Gaussian function to the onset of the WL in the APT data, see Fig.~\ref{figure_2}c~(top). The FWHM of this Gaussian function is 0.8~nm and thus can be considered as the instrument profile. This value compares favourably with the resolution obtained from laser-assisted evaporation of Si isotope superlattices.\cite{shimizu_atom_2009} An alternative technique, known as $z$-density correction, can also be used to improve interface sharpness in the depth direction.\cite{sauvage_solid_2001}

The deconvoluted APT data is shown in Fig.~\ref{figure_2}c~(bottom). The onset of the WL is now found to occur within $\approx0.35$~nm, which is well within one bilayer (0.565~nm), and the peak Indium fraction, $x$, has risen by $\approx2\%$, in agreement with the XSTM data. An exponentially decaying function in the same form as the input for the FE calculations was used to fit the deconvolved APT data. We find $a=0.189\pm0.004$ and $b=1.95\pm0.05$.

Comparing the two techniques we find an excellent match of the In segregation profiles. However, to arrive at this result a deconvolution of the APT data was necessary. In case no deconvolution function is known and $z$-density correction is not applicable, the APT's ability to image sharp interfaces is limited without better data reconstruction techniques.

\subsection{Quantum Dots}

Having shown, with the aid of FE calculations, that the two techniques give comparable results for the WLs we shall now consider the QDs. A total of 55 QDs, 43 by XSTM and 12 by APT, were characterized.

\begin{figure}
\includegraphics[width=8.6cm]{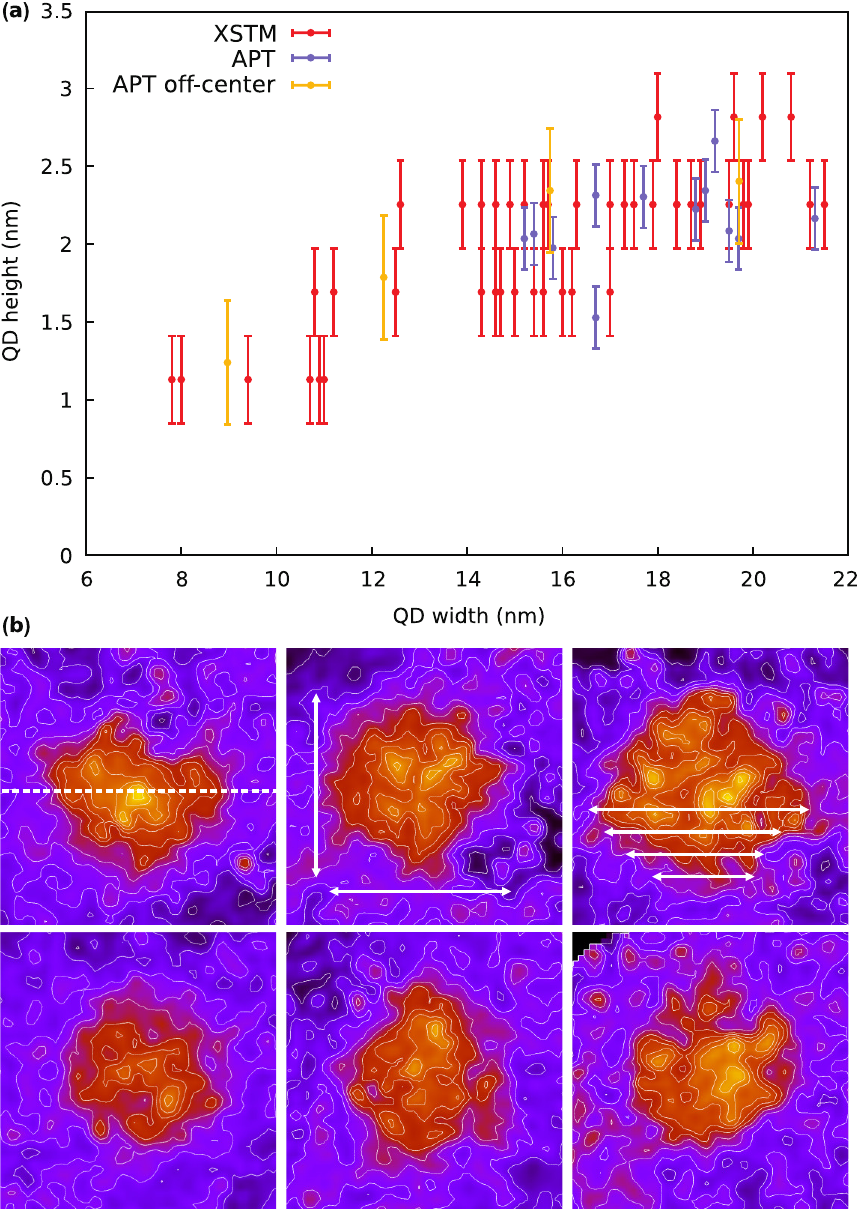}
\caption{QD height\,/\,width distribution. (a) The height and width of the QDs as measured with XSTM (red) was determined from topographic and current images. The APT height and width data (blue) is derived respectively from 1D composition profiles of a 10~nm diameter cylinder and cross-sectional contour plots using a 1~nm projection made through the centre of the QD. The yellow APT width data is derived from In concentration contour plots made at 2~nm intervals from the centre of a specific QD and the height data is extracted from the $x=25\%$ contour. (b) In concentration contour maps (25~nm\,$\times\,25$~nm) of six different QDs looking at the (001) surface. The images are orientated so the nominal longest axis is horizontal. The upper-left panel shows the centre position used to make the cross-sectional contour of Fig.~\ref{figure_4}c. The upper-centre panel illustrates how the width was estimated from cross-sectional contour plots. The arrows in the upper-right panel mark the length measured for the yellow points in (a). Three more QD footprints are depicted in the bottom panels. \label{figure_3}}
\end{figure}

In Fig.~\ref{figure_3}, the height and width of all observed QDs is plotted. The height of the QDs was determined from the XSTM data by counting the number of bilayers in the current images, resulting in the discrete nature of the plotted data. In APT the height of the QDs was determined from a 1D composition profile sampled by a 10~nm diameter cylinder going though the centre of the QD. The width is determined from the cross-sectional contour map with a 1~nm projection taken through what is estimated to be the longest part of the dot. An example of the position of the cross-section and example measurements of major and minor axes are shown in the upper-left and upper-centre panels of Fig.~\ref{figure_3}b, respectively. \footnote{See EPAPS Document No. [number will be inserted by publisher] for additional visualizations of the QD. For more information on EPAPS, see http://www.aip.org/pubservs/epaps.html.} The thickness of the highest QD as measured by XSTM is in agreement with the highest QD observed by APT (2.8~nm vs. 2.7~nm). At the lower end of the observed height distribution a discrepancy between the two techniques is observed; the thinnest QD found by XSTM is significantly lower than the thinnest QD observed by APT ($1.1$~nm vs. $1.5$~nm). This can be explained by the 2D nature of the XSTM technique. Due to the arbitrary position of the cleavage plane we have no \textit{a priori} way of knowing how a QD is cleaved. If a QD is cleaved near its edge it will appear thinner. The same holds for the width of the QDs. This results in a wide range (7--22~nm) of observed QD widths and in the trend of increasing QD height with QD width in the XSTM data seen in Fig.~\ref{figure_3}a.

In the fully 3D dimensional technique of APT the height and width of the QDs can be determined with less ambiguity. It is therefore not surprising that the height\,/\,width distribution of the QDs as measured by APT is clustered in the top right of Fig.~\ref{figure_3}a. The spread in the distribution reflects the non-uniformity of the growth process. By taking cross-sectional contour maps through non-central positions of the QD, as shown in the upper-right panel of Fig.~\ref{figure_3}b, we can emulate the effect of off-centre cleavage. The yellow points in Fig.~\ref{figure_3}a show the width and height, measured as the contour map is moved towards the edge of the QD. In contrast to the blue points, where a 1D profile was used to determine the height, in this case it is measured from the $x=25\%$ In concentration contour; this is not as accurate, but it gives more localized measurements.

It should also be noted here that there is a substantial error in the width of the QDs with both techniques. Due to the gradual transition of WL to QD it is hard to pinpoint where the QD starts. In an effort to increase the accuracy, a combination of current and topographic data was used to determine the width of the QDs with XSTM. In the case of the APT the WL profile is first subtracted from the data. However, with both techniques, an error of $\approx \pm 2$~nm remains in the width of the QDs. For reason of clarity these error bars are omitted from Fig.~\ref{figure_3}.

\begin{figure}
\includegraphics[width=8.6cm]{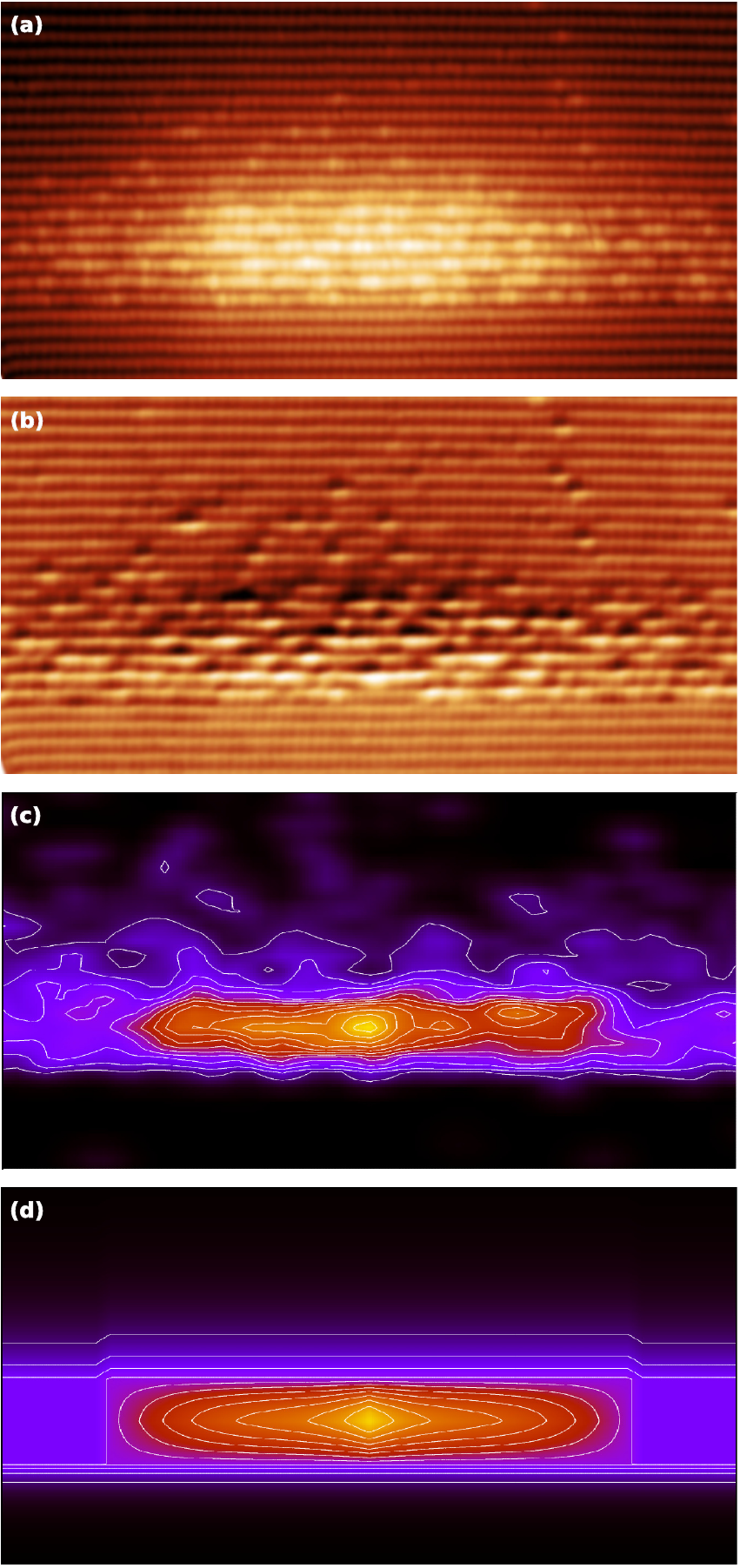}
\caption{Comparative views of two QDs. All images are to the same scale ($25$~nm\,$\times\,13$~nm).(a) Topographic and (b) local mean equalization filtered current XSTM image taken at $V=-3$~V, $I=40$~pA. The height was determined to be 2.8~nm. (c) contour plot of the In concentration of a 1~nm thick slice through the centre of a QD measured by APT. The 3D grid parameters used to create the 2D profile are a de-localization of 1.5~nm along the in-plane axes and 0.75~nm in the growth direction. The contour lines are every 5\%, ranging from 0\% to 65\%. The height of the QD as measured by APT was determined to be 2.3~nm. The APT data was used to construct a model (d) which serves as an input for the FE calculations. \label{figure_4}} 
\end{figure}

Figure~\ref{figure_4}a-b shows the topographic and the local mean equalization filtered current XSTM image of a cleaved QD. The height and width of this QD was determined to be 2.8~nm and 18.0~nm, respectively. As mentioned above, it is normally not known how the cleavage plane intersects the QDs with the XSTM technique. However, the statistics of Fig.~\ref{figure_3} show that this QD is one of the highest and widest that was observed. Therefore, we can reasonably assume that this particular QD was cleaved through its centre.\cite{Offermans2005a} Since not all the individual In atoms can be identified in the XSTM images, it is not possible to extract directly an In profile from Fig.~\ref{figure_4}a-b, and an indirect route via APT and FE calculations has to be taken. With APT the In profile of the QDs can readily be obtained. A contour plot of the In concentration of a slice through the centre of a QD that was observed by APT is shown in Fig.~\ref{figure_4}c. The QD and slice position is shown in the upper-left panel of Fig.~\ref{figure_3}b. Because this particular QD was situated near the middle of the APT specimen, and thus suffered least from reconstruction artifacts, it was chosen to link the two techniques of XSTM and APT. The height and width was determined to be 2.3~nm and 16.7~nm respectively, making this a typical QD. We would like to note here that a previous APT study has reported the peak In concentration in InAs QDs to be off-centred and towards the top of the QDs.\cite{Muller2008} This is not the case in our QDs where the peak In concentration is slightly inclined towards the top and laterally located in the centre. 

Conventionally, InAs/GaAs QDs are modelled by a disk or a truncated pyramid with increasing (from bottom to top) \cite{Bruls2002} or inverted-triangular shaped In profiles \cite{Liu2000}. Given the XSTM and APT data, such an approach cannot be followed with the current QDs. Figure~\ref{figure_3}b shows no strong evidence of faceting or a particular footprint for these QDs. As can be seen from the contour concentration maps the QDs merge into the WL, making edges indistinct, and each has a unique irregularity. Therefore, we consider it to be a reasonable first approximation to model these QDs as long thin spheroids with a circular footprint.

The size and shape of buried InAs/GaAs QDs is determined by a delicate interplay between driving and quenching of QD levelling during overgrowth \cite{Gong2004}. Depending on specifics of the growth conditions, the obtained QDs can be classified into two general types. The first class is characterized by a sharp interface between the QD and the surrounding matrix and is the result of a low growth temperature \cite{Gong2004} or a low growth rate \cite{Offermans2005,Bruls2002,Ulloa2009}. In case of a low growth temperature the segregation of In out of the WL is strongly suppressed, resulting in a localized WL. In contrast to this, if the first class of QD is formed by a low growth rate then a pronounced WL will be present.
The second class of QD is associated with a growth process involving a high growth temperature and a high growth rate. Under such growth conditions there is a strong In segregation (both vertically and laterally) during the overgrowth of the QDs. This results in a pronounced WL and softening of the interface of the QD with the surrounding matrix \cite{Offermans2005,Gong2004,Lita1999}.

\begin{figure}
\includegraphics[angle=0,width=8.6cm]{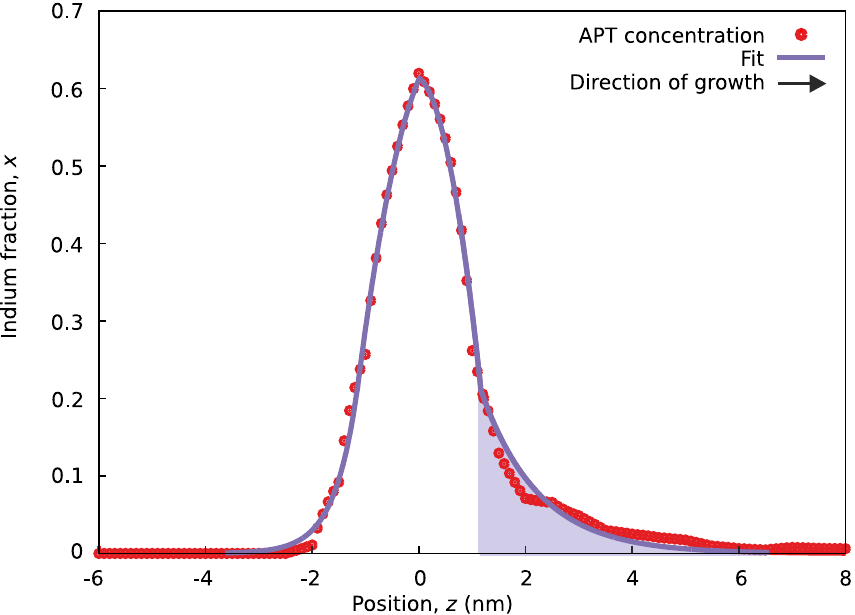}
\caption{1D In profile in the direction of growth through the centre of the QD shown in Fig.~\ref{figure_4}c. Analytical functions (solid lines) are fitted to the APT data. The In found above the QD in this profile is marked by the blue shaded area and amounts to 16\% of the total amount.}
\label{figure_5}
\end{figure}

From figure~\ref{figure_4}a-b we observe that the interfaces of our studied QDs are not sharp and a pronounced WL is present. It is, therefore, clear that these QDs do not belong to the first class and are instead examples of the second class of QD. We would like to note that it can be problematic to distinguish between these two classes in most XSTM and TEM studies of InAs/GaAs QDs, thus particular care must be taken when trying to model the QDs. To aid in the construction of a model of our QDs two 1D In profiles were extracted from the APT data shown in Fig.~\ref{figure_4}c: one through its centre along in the growth direction, shown in Fig.~\ref{figure_5}, and one laterally through the centre, perpendicular to the growth direction (not shown). These data are fitted with analytical functions and scaled to match the width and height of the QD observed by XSTM depicted in Fig.~\ref{figure_4}a-b. The analytical expressions were then extended in order to generate a fully 3D model of the QD.\footnote{See EPAPS Document No. [number will be inserted by publisher] for numeric details of the QD model. For more information on EPAPS, see http://www.aip.org/pubservs/epaps.html.} The result is shown in Fig.~\ref{figure_4}d.

Besides the In profile in the core of the QD, two other regions proved crucial in modelling the QD. Firstly, the exponential decay of the In concentration in the direction of growth above the QD. From the 1D In profile taken through the centre of the QD, we determined that this amounts to 16\% of the total number of In atoms present; see the marked area in Fig.~\ref{figure_5}. Note that in the XSTM images of Fig.~\ref{figure_4}a-b, only a few individual In atoms are visible above the QD and that an exponential decay cannot be distinguished. Because the area above the QDs measured by XSTM is very limited a meaningful statistical analysis is not practicable. Consequently, the In in this region are often overlooked.
However, the APT data reveals that, together, these In atoms actually form a significant part of the Indium profile and should thus be included in the model. The second region that is crucial in modelling our QDs is the gradual transition of the QD into the WL. The XSTM and APT images of Fig.~\ref{figure_4}a-c show that the In concentration at the sides of the QD is more substantial than that of the WL which the profiled in the previous section. We choose to model this feature as a region of $x=15\%$ In fraction which extends well beyond the QD and eventually merges with the WL (not shown in Fig.~\ref{figure_4}d).

\begin{figure}
\includegraphics[angle=0,width=8.6cm]{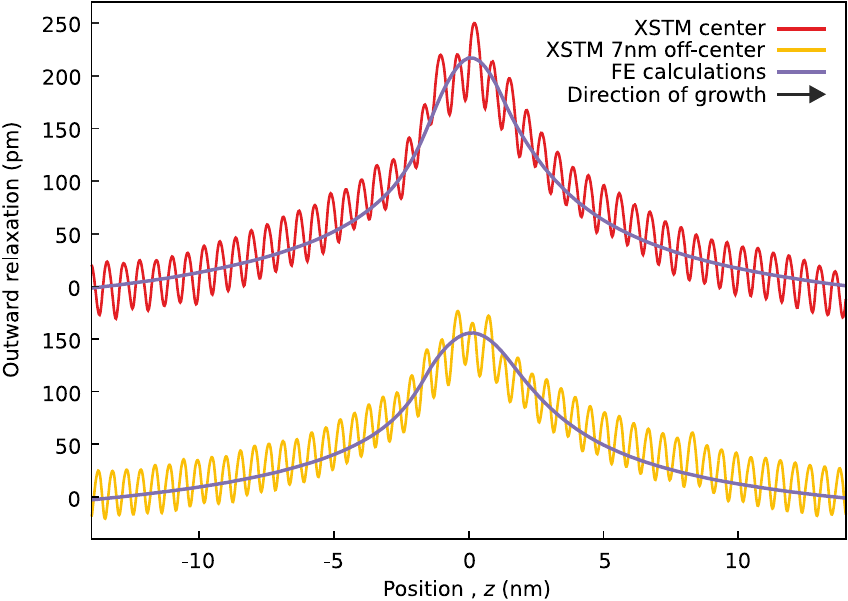}
\caption{Outward relaxation of the cleaved surface of the QD shown in Fig.~\ref{figure_4}a-b. The outward relaxation measured by XSTM through the centre (red line) and 7~nm off-centre (yellow line) of the QD is compared with the results of the FE calculations (blue lines).}
\label{figure_6}
\end{figure}

In Fig.~\ref{figure_6}, the outward relaxation of the cleaved surface across the centre and 7~nm off-centre of the QD, as measured by XSTM, is plotted together with the result of the FE calculations. A close match between the measured and the calculated outward relaxation is observed. Here it should be noted that different input models, e.g. the models with a linear In profile mentioned above, can yield a similar match. However, such models would not resemble the APT data. From this we conclude that care is required when constructing QD models based on XSTM measurements alone. Recently, it has been shown \cite{Mlinar2009} that PL measurements in combination with extensive theoretical modelling and XSTM measurements had to be combined to yield a realistic QD model. However, the subtleties of the decaying In concentration above the QD and the extension of the QD into the WL might be overlooked in such an approach, and thus some ambiguity still remains. In our view, the abilities of APT in providing a 3D In profile are at the moment unique, and necessary to construct an valid 3D QD model.

\section{Conclusion}

We have used XSTM and APT to image InAs/GaAs QD layers in a single sample. We have found there is good agreement between the segregation profiles of the WL obtained by both techniques. To arrive at this result it was necessary to deconvolve the APT data with an instrument profile. The height and length of the QDs determined by both techniques is also comparable, both in the case where cross-sections are taken through the centre of the QD and also where they are taken off-centre. This highlights the versatility of APT whereby the 3D data can be processed in a variety of ways to show a variety of details. Exploiting this advantage, we have used the In profile measured by APT to make a 3D model of a typical dot. This model is in agreement with the outward relaxation measured by XSTM, and we therefore consider this to be a unique solution. This analysis method highlights some structural features of the QDs that were undetected or neglected in previous measurements. The juxtaposition of the two techniques shows the unique benefits and capabilities of each one. Thanks to their very different natures, by using XSTM and APT together we can analyse semiconductor nanostructures at a level of detail that has been previously unobtainable.

\section*{Acknowledgements}

We would like to thank: STW-VICI for their financial support under Grant No. 6631; N. Arai and H. Uchida, of Toshiba Nanoanalysis, for the FIB fabrication; and M. M\"uller for the fruitful discussions.


%

\clearpage
\onecolumngrid
\appendix*

\section{Auxiliary Material}

\begin{center}
\begin{figure}[h]
\includegraphics[height=9.15cm]{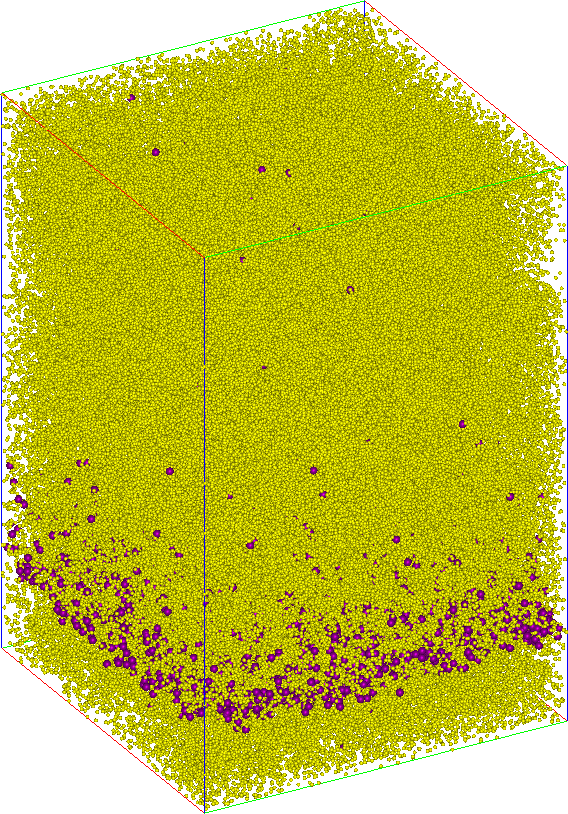}
~
\includegraphics[height=9.15cm]{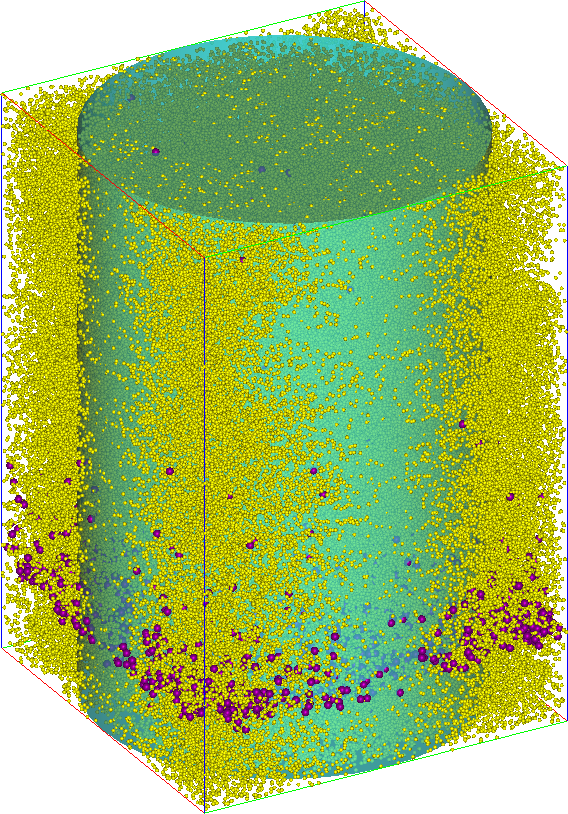}
\caption{Example atom probe data showing a $20$~nm\,$\times\,20$~nm\,$\times\,30$~nm volume through a section of the (In,Ga)As WL. For clarity, only the group-III elements are shown. The In atoms are represented by indigo spheres with a radius of 0.2~nm and the Ga atoms are represented by yellow spheres with a radius of 0.1~nm. The right-hand figure shows a cylinder with a 20~nm diameter, as was used to sample the 1D composition profiles of the WL for Fig.~2b in the main article.}
\end{figure}
\end{center}

\begin{center}
\begin{figure}[h]
\includegraphics[width=8.5cm]{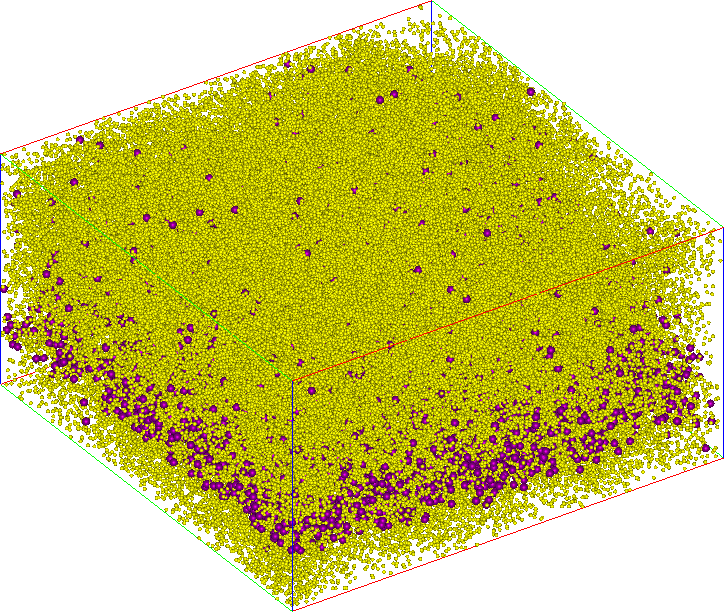}
\includegraphics[width=8.5cm]{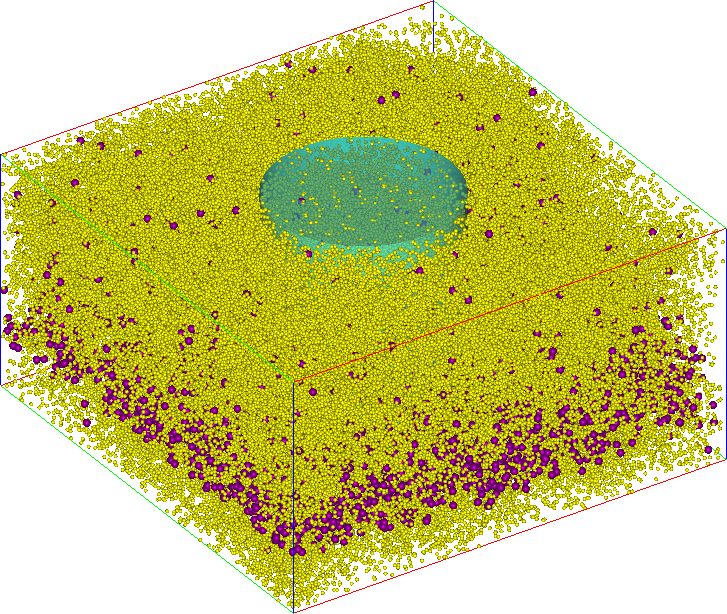}
\caption{A $25$~nm\,$\times\,25$~nm\,$\times\,13$~nm atom map of a volume of (In,Ga)As containing a QD. For clarity, only the group-III elements are shown.
Shown in the right-hand figure is a cylinder with a 10~nm diameter, as was used to sample the 1D composition profile in order to estimate the QD height for Fig.~3a in the main article.}
\label{EPASP_figure_QD}
\end{figure}
\end{center}

\begin{center}
\begin{figure}[h]
\includegraphics[width=8.5cm]{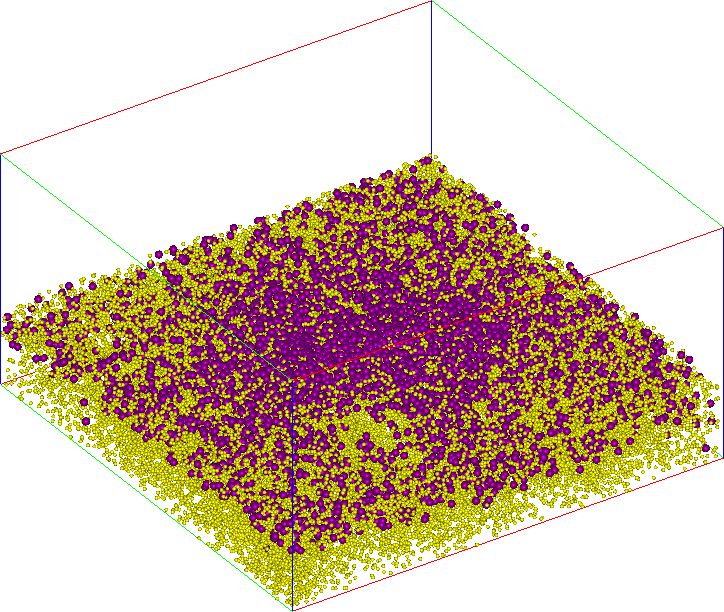}
\includegraphics[width=8.5cm]{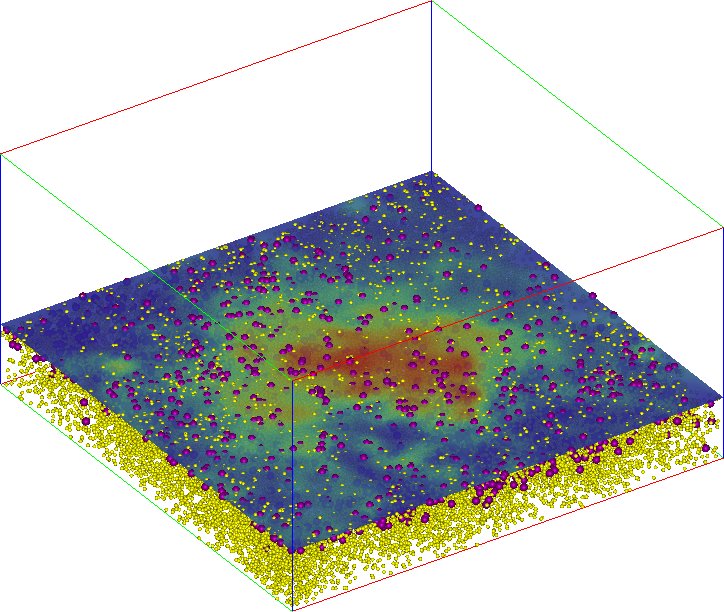}
\caption{A cross-section of the data in Fig.~\ref{EPASP_figure_QD}, taken normal to the growth direction, revealing the footprint of the QD. In the right image a 2D conentraction profile, sampled from a volume with a 1~nm projection, has been superimposed on this area. This data can be seen in the top-left panel of Fig. 3b in the main article.}
\end{figure}
\end{center}

\begin{center}
\begin{figure}[h]
\includegraphics[width=8.5cm]{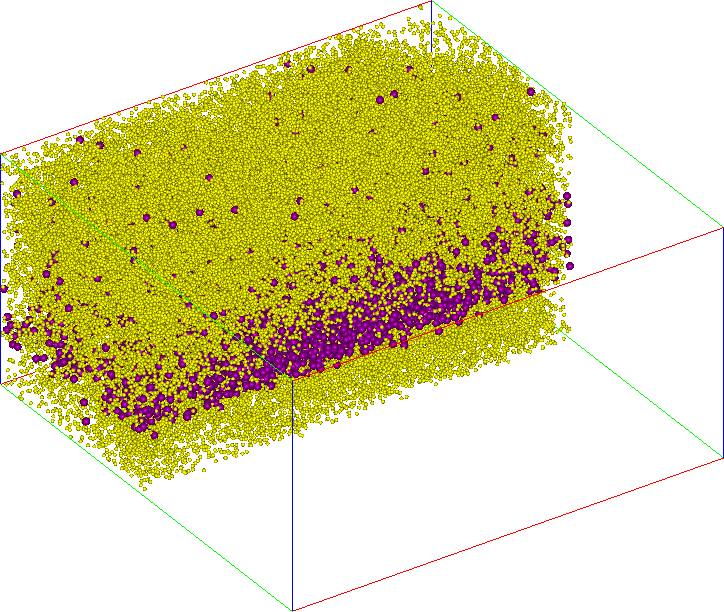}
\includegraphics[width=8.5cm]{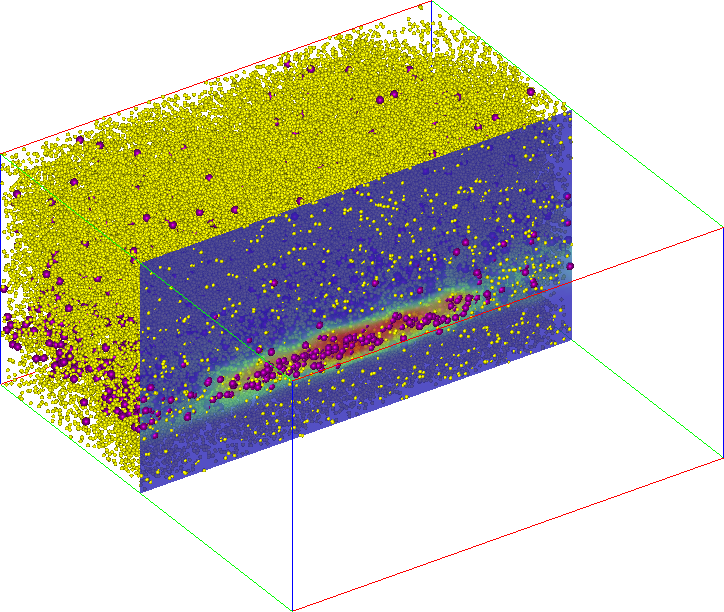}
\caption{A cross-section of the data in Fig.~\ref{EPASP_figure_QD}, taken through the centre of the volume, revealing the core of the QD. In the right image a 2D conentraction profile, sampled from a volume with a 1~nm projection, has been superimposed on this area. This data can be seen in Fig.~4c in the main article.}
\end{figure}
\end{center}

\begin{center}
\begin{figure}[h]
\includegraphics[width=17.2cm]{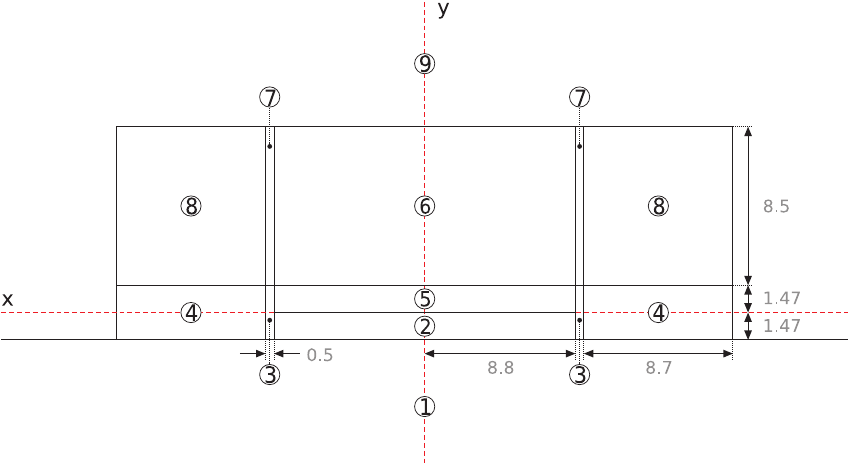}
\caption{Details of the QD model used as input for the FE calculations. The dimensions of the total box enclosing the QD are 100~nm$^3$. Dimension are in nm. Different areas are marked. The contour plot of this model is shown in Fig.~4d of the main article.}
\label{EPASP_figure_1}
\end{figure}
\end{center}

\begin{center}
\begin{table}[h]
\caption{Analytical expressions for the In fraction for the areas indicated in Fig.~\ref{EPASP_figure_1}. The coordinate system is Cartesian.}
	\begin{tabular}{|p{1cm}|p{12cm}|p{4cm}|}
		\hline
		Area & Expression & description \\
		\hline		
		1 & $0.21\exp((z+1.47)/0.43)$ & transitional region \\ 
		2 & $((0.0003(x^2+y^2)^2-0.0088(x^2+y^2)^{\frac{3}{2}}+0.0269(x^2+y^2)-0.0645(x^2+y^2)^{\frac{1}{2}}+1)(-0.1023z^2+0.1243z+0.4016)+0.21$ & lower part QD\\
		3 & $0.22-0.14((x^2+y^2)^{\frac{1}{2}}-8.8)$ & transitional region \\
		4 & $0.151$ & transition WL-QD\\
		5 & $((0.0003(x^2+y^2)^2-0.0088(x^2+y^2)^{\frac{3}{2}}+0.0269(x^2+y^2)-0.0645(x^2+y^2)^{\frac{1}{2}}+1)(-0.1638z^2-0.0317z+0.4032)+0.21$ & upper part QD\\
		6 & $0.20\exp((-z+1.53)/1.08)$ & indium rich region above QD \\
		7 & $(0.20-0.10((x^2+y^2)^\frac{1}{2}-8.8)\exp((-z+1.53)/1.08)$ & transitional region \\
		8 & $0.15\exp((-z+1.53)/1.08$ & transitional region\\
		9 & $0.19\exp((-z-1.47)/2))$ & wetting layer\\
		\hline
	\end{tabular}
	\label{tab1}
\end{table}
\end{center}

\end{document}